\documentclass{SciPost}

\hypersetup{
    colorlinks,
    linkcolor={red!50!black},
    citecolor={blue!50!black},
    urlcolor={blue!80!black}
}

\usepackage[bitstream-charter]{mathdesign}
\DeclareSymbolFont{usualmathcal}{OMS}{cmsy}{m}{n}
\DeclareSymbolFontAlphabet{\mathcal}{usualmathcal}

\fancypagestyle{SPstyle}{
\fancyhf{}
\lhead{\raisebox{-1.5mm}[0pt][0pt]{\href{https://scipost.org}{\includegraphics[width=20mm]{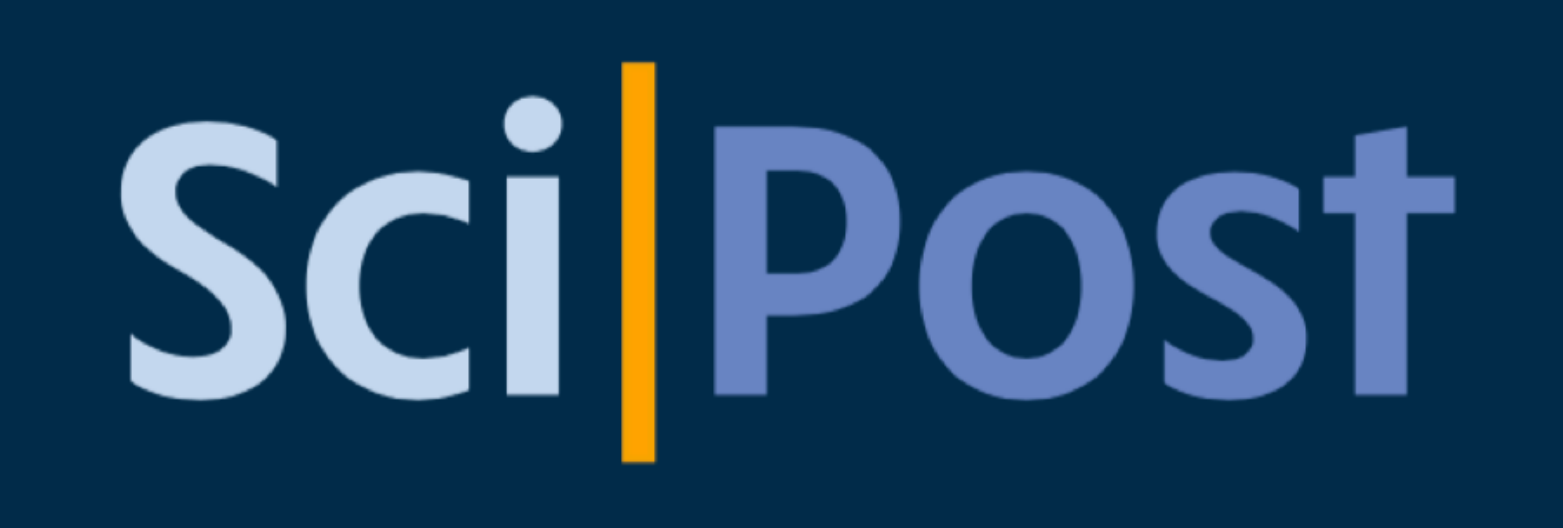}}}}
\rhead{\small \href{https://scipost.org/SciPostPhys.15.5.205}{SciPost Phys. 15, 205 (2023)}}

\fancyfoot[C]{\textbf{\thepage}}
}

\newcommand\abstr[1]{{\boldmath\textbf{#1}}}

\usepackage{booktabs}
\usepackage{xspace}
\usepackage{dcolumn}
\usepackage
[subrefformat=parens,position=top,skip=-15pt,margin=15pt,justification=justified,singlelinecheck=false]
{subcaption}
\usepackage{array,multirow}
\usepackage{braket}
\usepackage{makecell}
\usepackage{dsfont}
\newcommand{\id}{\mathds{1}}

\newcommand\matthree[9]{%
  \begin{pmatrix}
  #1 & #2 & #3 \\ #4 & #5 & #6 \\ #7 & #8 & #9
  \end{pmatrix}%
}
\makeatletter
\g@addto@macro\bfseries{\boldmath}
\makeatother

\begin{document}

\pagestyle{SPstyle}

\begin{center}{\Large \textbf{\color{scipostdeepblue}{
Quantum simulation of colour in perturbative\\ quantum chromodynamics\\
}}}\end{center}

\begin{center}
\textbf{
Herschel A. Chawdhry\textsuperscript{1$\star$} and
Mathieu Pellen\textsuperscript{2$\dagger$}
}
\end{center}

\begin{center}
{\bf 1} Rudolf Peierls Centre for Theoretical Physics, University of Oxford,\\ Clarendon Laboratory, Parks Road, Oxford, OX1 3PU, United Kingdom
\\
{\bf 2} Albert-Ludwigs-Universit\"at Freiburg, Physikalisches Institut,\\ Hermann-Herder-Stra\ss e 3, D-79104 Freiburg, Germany
\\[\baselineskip]
$\star$ \href{mailto:herschel.chawdhry@physics.ox.ac.uk}{\small herschel.chawdhry@physics.ox.ac.uk}\,,\quad
$\dagger$ \href{mailto:mathieu.pellen@physik.uni-freiburg.de}{\small mathieu.pellen@physik.uni-freiburg.de}
\end{center}

\section*{\color{scipostdeepblue}{Abstract}}
\abstr{%
Quantum computers are expected to give major speed-ups for the simulation of quantum systems. In this work, we present quantum gates that simulate the colour part of the interactions of quarks and gluons in perturbative quantum chromodynamics (QCD). As a first application, we implement these circuits on a simulated noiseless quantum computer and use them to calculate colour factors for various examples of Feynman diagrams. This work constitutes a first key step towards a quantum simulation of generic scattering processes in perturbative QCD.
}

\begin{center}
\begin{tabular}{lr}
\begin{minipage}{0.6\textwidth}
\raisebox{-1mm}[0pt][0pt]{\includegraphics[width=12mm]{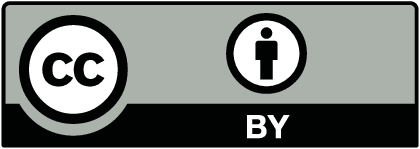}}
{\small Copyright H. A. Chawdhry and M. Pellen. \newline
This work is licensed under the Creative Commons \newline
\href{http://creativecommons.org/licenses/by/4.0/}{Attribution 4.0 International License}. \newline
Published by the SciPost Foundation.
}
\end{minipage}
&
\begin{minipage}{0.4\textwidth}
    \noindent\begin{minipage}{0.68\textwidth}
    {\small Received 17-03-2023 \newline Accepted 10-11-2023 \newline Published 24-11-2023}
    \end{minipage}
    \begin{minipage}{0.25\textwidth}
    \begin{center}
    \href{https://crossmark.crossref.org/dialog/?doi=10.21468/SciPostPhys.15.5.205&amp;domain=pdf&amp;date_stamp=2023-11-24}{\includegraphics[width=7mm]{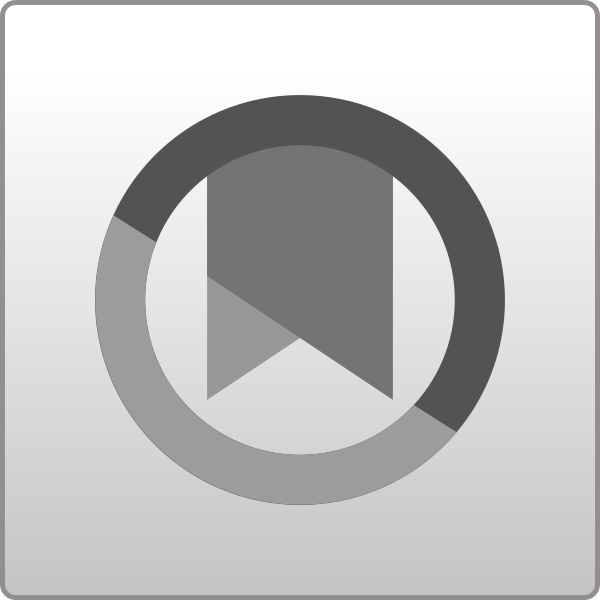}}\\
    \tiny{Check for}\\
    \tiny{updates}
    \end{center}
    \end{minipage}
    \\\\
    \small{\doi{10.21468/SciPostPhys.15.5.205}}
\end{minipage}
\end{tabular}
\end{center}

\vspace{10pt}
\noindent\rule{\textwidth}{1pt}
\tableofcontents
\noindent\rule{\textwidth}{1pt}
\vspace{10pt}

\section{Introduction}

Quantum computing is of widespread interest because it offers exponential or polynomial speedups for a variety of problems ranging from prime factorisation~\cite{Shor.365700} to unstructured searching~\cite{10.1145/237814.237866}.
A natural use of quantum computers is the simulation of other quantum systems, with well-known applications in computational chemistry~\cite{ChemistryRev,RevModPhys92015003} and condensed matter physics~\cite{RevModPhys.86.153,qute.201900052}.

Recent years have seen proposed applications of quantum computers to lattice-based Quantum Field Theory (QFT) simulations~(see Refs.~\cite{Klco:2021lap,Bauer:2022hpo} and references therein) including the simulation of Quantum Chromodynamics (QCD), the theory describing the fundamental interactions of quarks and gluons.
Lattice QCD is well-suited to studying the lower-energy (sub-GeV) behaviour of QCD, but the rapid increase in computational cost with lattice size makes lattice QCD exceedingly challenging to use for simulating collisions at the shortest length scales probed in high-energy colliders such as the Large Hadron Collider (LHC).
At these energies, the QCD coupling constant $\alpha_s$ becomes small, and so perturbative calculations become the method of choice.

The use of quantum computers to simulate hard scattering processes in perturbative QCD has largely remained unexplored to date.
This may be in part because the aims of perturbative QFT calculations differ from the aims of most quantum simulations: most quantum simulations (including lattice QCD) aim to take a known Hamiltonian and use it to perform the (unitary) evolution of a quantum system, whereas perturbative QFT calculations aim to calculate the (Hermitian, but not unitary) transition matrix describing the scattering of specified external states and hence study the production or decay of elementary particles.

A method to simulate generic perturbative QCD processes on a quantum computer is still missing but is desirable for several reasons.
Firstly, perturbative QCD calculations require the evaluation---and quantum coherent combination---of contributions from many unobservable intermediate states, which makes such calculations natural candidates to benefit from the ability of quantum computers to manipulate superpositions of quantum states.
Secondly, this ability also means quantum simulations could be well-suited to performing calculations with full interference effects for processes with high-multiplicity final states.
Thirdly, a quantum simulation of generic perturbative QCD processes could improve the speed and precision of perturbative QCD predictions by exploiting speedups provided by known quantum algorithms such as quantum amplitude estimation~\cite{Brassard:2000,Grinko:2019,Suzuki:2019,Nakaji:2020}.

The object of this article is to take steps towards using quantum computers to simulate generic perturbative QCD processes.
Calculations in perturbative QCD can be performed by summing contributions from Feynman diagrams. Each contribution can be factorised into a colour part and a kinematic part.
The colour part is simpler to compute than the kinematic part, and indeed there exist efficient programs~\cite{Bern:1990ux,Hakkinen:1996bb,Kanaki:2000ey,Papadopoulos:2005ky,Sjodahl:2012nk,Actis:2012qn} for calculating colour factors on a classical computer.
Nonetheless, the colour part still presents some of the generic challenges of simulating perturbative QCD processes on a quantum computer.
For example, the quantum gates that form a quantum computer must always be unitary whereas the Feynman rules (colour and kinematic parts alike) describing components of a Feynman diagram are not generally unitary.
This means the colour parts provide a useful simplified setup with which to begin developing a framework for the quantum computation of Feynman diagrams, and they will therefore be the focus of the present work.

The main results in this article are two quantum gates, $Q$ and $G$, which represent the colour part of the Feynman rules describing the quark-gluon and the triple-gluon interaction vertices, respectively.
To implement these gates, we introduce the new concept of a \emph{unitarisation register} $\mathcal{U}$, which enables the simulation of the non-unitary interactions of quarks and gluons.\footnote{While the use of ancillary qubits to enforce unitarity is by now well established with methods such as \emph{block encoding}~\cite{Lloyd,Kimmel,vanapeldoorn,QuantumSingularValue} or \emph{qubitisation}~\cite{Low2016HamiltonianSB}, to the best of our knowledge our implementation is original and has the advantage of allowing multiple independent non-unitary operations to be carried out sequentially while only requiring a small number of ancilla qubits, as will be explained in sec.~\ref{sec:unitarisation_register}.}
As an example of the use of $Q$ and $G$, we use the {\sc Qiskit} \cite{Qiskit} quantum-computing framework to build quantum circuits that calculate the colour factors of various Feynman diagrams.
This is by no means the only use case for $Q$ and $G$; they could also be used, for example, to simulate emissions, absorption, and exchanges of quarks and gluons, either in a scattering amplitude or in a parton shower.
These quantum circuits could also form a component of a quantum Monte Carlo program which, as highlighted in Ref.~\cite{Agliardi:2022ghn}, would offer a quadratic speedup for the calculation of cross sections.

Finally, let us mention that this work forms part of a broader exploration of quantum-technology applications in high-energy physics (see \emph{e.g.}\ existing reviews~\cite{Klco:2021lap,Bauer:2022hpo,Gray:2022fou,Delgado:2022tpc} and references therein).
While most of that exploration has focused on the experimental side of high-energy physics, the last few years have also seen the emergence of applications on various topics in high-energy theory.
These range from parton distribution functions (PDFs)~\cite{Perez-Salinas:2020nem,Li:2021kcs} to amplitudes~\cite{Bepari:2020xqi,Ramirez-Uribe:2021ubp,Cervera-Lierta:2017tdt,Fedida:2022izl,Clemente:2022nll}, effective field theory~\cite{Bauer:2021gup}, cross-section computations~\cite{Agliardi:2022ghn}, parton showers~\cite{Bepari:2020xqi,Bauer:2019qxa,Bepari:2021kwv,Gustafson:2022dsq}, and event generation~\cite{Gustafson:2022dsq,Bravo-Prieto:2021ehz,Kiss:2022pjw}.

This article is organised as follows:
sec.~\ref{sec:methods} begins with a high-level overview of the use of our quantum circuits to calculate the colour factor for a simple Feynman diagram~(sec.~\ref{sec:colour_traces}).
The rest of sec.~\ref{sec:methods} provides details of the methods and algorithms employed.
In particular, sec.~\ref{sec:unitarisation_register} explains the functioning of the unitarisation register mentioned above, and sec.~\ref{sec:gates} presents the quantum circuits implementing quark and gluon interactions.
In sec.~\ref{sec:results}, we generalise our methods to simulate more complicated processes and validate this by using a simulated noiseless quantum computer to calculate colour factors for various Feynman diagrams.
Finally, sec.~\ref{sec:conclusion} contains a summary of our findings and concluding remarks.
In Appendix~\ref{appendix:misc_gates}, a few miscellaneous quantum gates related to the calculation of traces are presented.

\section{Methods}
\label{sec:methods}

\subsection{Illustrative example}
\label{sec:colour_traces}

\begin{figure}[]
\center
        \begin{subfigure}{0.35\textwidth}
                 \includegraphics[width=\textwidth]{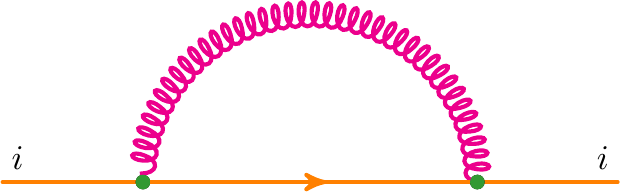}
        \end{subfigure}
\quad
        \begin{subfigure}{0.60\textwidth}
                 \includegraphics[width=\textwidth]{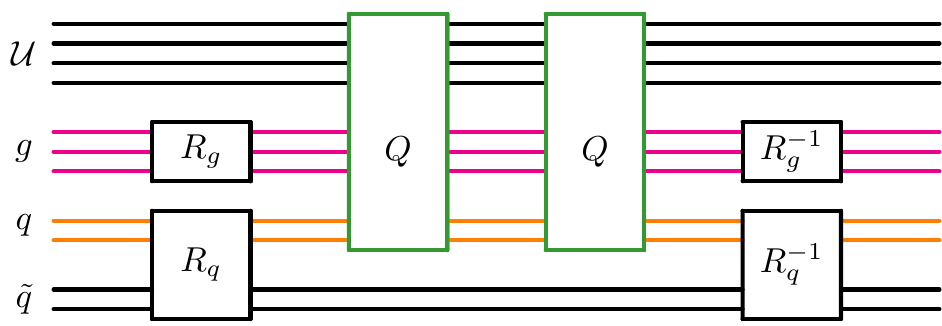}
        \end{subfigure}
        \caption{\label{fig:quarkselfenergy}%
                Example Feynman diagram (left) and a graphical representation of its corresponding circuit (right).}
\end{figure}

The main results in this article are two quantum gates, $Q$ and $G$, which simulate, respectively, the colour factors $T^a_{ij}$ of the quark-gluon vertex and $f^{abc}$ of the triple-gluon vertex.
We defer a description of the explicit construction of $Q$ and $G$ to sec.~\ref{sec:gates}.
As we will see in sec.~\ref{sec:results}, these gates can be used to calculate the colour factor of any Feynman diagram.
In the present section, we will illustrate how our method works by applying it to calculate the colour factor of the simple example Feynman diagram shown on the left-hand side of fig.~\ref{fig:quarkselfenergy}.

In general, any Feynman diagrams involving quark-gluon or gluon-gluon interactions will carry colour information from the SU(3) symmetry group of QCD.
When squaring the diagrams to obtain the cross section, the colour algebra has to be carried out, resulting in so-called colour factors.
The latter contain a generator $T^a$ for each quark-gluon vertex and a structure constant $f^{abc}$ for each triple-gluon vertex.\footnote{More references and information on colour algebra can be found in Refs.~\cite{Dixon:1996wi,Mangano:1998fk}.}
In the present example, the colour factor reads
\begin{equation}\label{eq:example_trace_analytic_result}
\mathcal{C} = \sum_{
\substack{
a \in \{1, ..., 8\}\\
i,j,k \in \{1,2,3\}
}
} T^a_{ij} T^a_{jk} \delta_{ik}\,,
\end{equation}
where
\begin{equation}\label{eq:T_is_half_lambda}
T^a = \frac{1}{2} \lambda^a\,,
\end{equation}
and $\lambda^a$ are the Gell-Mann matrices:
\begin{equation}\label{eq:lambas}
\begin{gathered}
\lambda^1 = \matthree {0}{1}{0}{1}{0}{0}{0}{0}{0}\,,\quad
\lambda^2 = \matthree {0}{-i}{0}{i}{0}{0}{0}{0}{0}\,,\quad
\lambda^3 = \matthree {1}{0}{0}{0}{-1}{0}{0}{0}{0},\\[1ex]
\lambda^4 = \matthree {0}{0}{1}{0}{0}{0}{1}{0}{0}\,,\quad
\lambda^5 = \matthree {0}{0}{-i}{0}{0}{0}{i}{0}{0}\,,\quad
\lambda^6 = \matthree {0}{0}{0}{0}{0}{1}{0}{1}{0},\\[1ex]
\lambda^7 = \matthree {0}{0}{0}{0}{0}{-i}{0}{i}{0}\,,\quad
\lambda^8 = \frac{1}{\sqrt{3}} \matthree {1}{0}{0}{0}{1}{0}{0}{0}{-2} \,.
\end{gathered}
\end{equation}
This colour factor can be computed using the quantum circuit shown on the right-hand side of fig.~\ref{fig:quarkselfenergy}, whose workings will now be explained.
The circuit contains several qubits, which we combine into groups called registers. Each register $r$ is initially in some reference state $\ket{\Omega}_r$.\footnote{In practice, in this work we always choose $\ket{\Omega}_r$ to be the state where each qubit of $r$ is in the state $\ket{0}$.}
In this particular example, there is a gluon register, labelled $g$, and a pair of quark registers, labelled $q$ and $\tilde{q}$. In general there will be a gluon register for each gluon in a Feynman diagram, and a pair of quark registers for each quark line in the diagram. We will see that the state of the $q$ register is altered by the simulated emission and absorption of gluons, while the $\tilde{q}$ register is left unaffected and serves only to help implement the $\delta_{ik}$ term in eq.~\eqref{eq:example_trace_analytic_result}.
Each quark register is made of 2 qubits, with the states $\ket{00}, \ket{01}, \ket{10}$ representing the $N_c = 3$ quark colours $\ket{1}, \ket{2}, \ket{3}$, while the state $\ket{11}$ is unused.
The gluon register is composed of 3 qubits, whose 8 states $\ket{000}, \ket{001}, \ldots, \ket{111}$ represent the $N_{\rm c}^2 -1 = 8$ colours $\ket{1}, \ket{2}, \ldots, \ket{8}$ of the gluon. There is also a unitarisation register, labelled $\mathcal{U}$, whose purpose will be explained in section~\ref{sec:unitarisation_register}.
The initial state of the circuit is thus $\ket{\Omega}_g\ket{\Omega}_q\ket{\Omega}_{\tilde{q}}\ket{\Omega}_\mathcal{U}$.

First, a gate $R_g$ is applied to the gluon register to put it into an equal superposition of colour states. A detailed definition of $R_g$ will be given in Appendix~\ref{appendix:misc_gates} and its effect reads as follows:
\begin{equation}
\label{eq:Rg_effect}
R_g\ket{\Omega}_g = \sum_{a=1}^8 \frac{1}{\sqrt{8}} \ket{a}_g\,.
\end{equation}
The gate $R_q$ (also to be defined in Appendix~\ref{appendix:misc_gates}) is now applied to the quark registers with the following effect:
\begin{equation}\label{eq:Rq_effect}
R_q \ket{\Omega}_q \ket{\Omega}_{\tilde{q}} = \sum_{k=1}^3 \frac{1}{\sqrt{3}} \ket{k}_q\ket{k}_{\tilde{q}}\,.
\end{equation}
Thus, after applying the $R_g$ and $R_q$ gates, the quantum computer is in the state
\begin{equation}
\frac{1}{\sqrt{24}} \sum_{a=1}^8 \sum_{k=1}^3 \ket{a}_g\ket{k}_q\ket{k}_{\tilde{q}}\ket{\Omega}_\mathcal{U}\,.
\end{equation}

Next, the $Q$ gate is applied to the $g$ and $q$ registers. $Q$ represents the quark-gluon interaction and is designed (see sec.~\ref{sec:quark_gate}) such that for a gluon colour basis state $\ket{a}_g$ and quark colour basis state $\ket{k}_q$, where  $a \in \{1, \ldots, 8\}$ and $k \in \{1,2,3\}$, the following equation holds:
\begin{equation}\label{eq:Tij_braket}
Q\ket{a}_g\ket{k}_q\ket{\Omega}_\mathcal{U} = \sum_{j=1}^3 T^a_{jk} \ket{a}_g\ket{j}_q\ket{\Omega}_\mathcal{U} + \left(\textrm{terms orthogonal to } \ket{\Omega}_\mathcal{U} \right) \,.
\end{equation}
Note that $Q$ does not act on the $\tilde{q}$ register.
The linearity of quantum gates ensures that $Q$ models the quark-gluon interaction correctly even if the quark or gluon registers are in superpositions of colour basis states or are entangled with other registers. Thus, after applying the $Q$ gate once, the quantum computer is in the state
\begin{equation}
\frac{1}{\sqrt{24}} \sum_{
\substack{
a \in \{1,\ldots,8\}\\
j,k \in \{1,2,3\}
}
} T^a_{jk} \ket{a}_g\ket{j}_q\ket{k}_{\tilde{q}}\ket{\Omega}_\mathcal{U} + \left(\textrm{terms orthogonal to } \ket{\Omega}_\mathcal{U} \right) \,.
\end{equation}
The $Q$ gate is now applied a second time to the $g$ and $q$ registers, to simulate the second quark-gluon vertex.
This puts the quantum computer into the state
\begin{equation}\label{eq:example_circuit_after_second_Q_gate}
\frac{1}{\sqrt{24}} \sum_{
\substack{
a \in \{1,\ldots,8\}\\
i,j,k \in \{1,2,3\}
}
} T^a_{ij}T^a_{jk} \ket{a}_g\ket{i}_q\ket{k}_{\tilde{q}}\ket{\Omega}_\mathcal{U} + \left(\textrm{terms orthogonal to } \ket{\Omega}_\mathcal{U} \right) \,.
\end{equation}

Since $R_g$ is unitary, one can see by inverting eq.~(\ref{eq:Rg_effect}) that $R_g^{-1}$ acting on any state $\sum_{a=1}^8 c_a \ket{a}_g$ of the gluon register would produce the state
\begin{equation}
R_g^{-1} \sum_{a=1}^8 c_a \ket{a}_g = \left(\frac{1}{\sqrt{8}} \sum_{a=1}^8 c_a\right)\ket{\Omega}_g + \left(\textrm{terms orthogonal to } \ket{\Omega}_g \right) \,.
\end{equation}
Similarly, it can be seen by inverting eq.~(\ref{eq:Rq_effect}) that $R_q^{-1}$ acting on any state $\sum_{i,k\in\{1,2,3\}} c_{ik} \ket{i}_q \ket{k}_{\tilde{q}}$ of the $q$ and $\tilde{q}$ registers would produce the state
\begin{equation}
R_q^{-1} \sum_{i,k\in\{1,2,3\}} c_{ik} \ket{i}_q \ket{k}_{\tilde{q}} = \left(\frac{1}{\sqrt{3}} \sum_{i=1}^3 c_{ii} \right)\ket{\Omega}_q \ket{\Omega}_{\tilde{q}} + \left(\textrm{terms orthogonal to } \ket{\Omega}_q\ket{\Omega}_{\tilde{q}} \right) \,.
\end{equation}
Therefore, taking the state produced in eq.~\eqref{eq:example_circuit_after_second_Q_gate} and applying the gates $R_g^{-1}$ and $R_q^{-1}$, the state of the quantum circuit becomes
\begin{equation}\label{eq:example_trace_final_state}
\frac{1}{24}\left(
\sum_{
\substack{
a \in \{1, ..., 8\}\\
i,j \in \{1,2,3\}
}
} T^a_{ij} T^a_{ji}
\right)
\ket{\Omega}_g\ket{\Omega}_q\ket{\Omega}_{\tilde{q}}\ket{\Omega}_\mathcal{U} + \left(\textrm{terms orthogonal to } \ket{\Omega}_g\ket{\Omega}_q\ket{\Omega}_{\tilde{q}}\ket{\Omega}_\mathcal{U} \right) \,.
\end{equation}
Thus, in the final quantum state~\eqref{eq:example_trace_final_state} of the quantum circuit, the colour trace~\eqref{eq:example_trace_analytic_result} to be computed is found encoded in the coefficient of the reference state $\ket{\Omega}_g\ket{\Omega}_q\ket{\Omega}_{\tilde{q}}\ket{\Omega}_\mathcal{U}$.
As we will now discuss, this circuitry can either be used as part of a higher-level algorithm, or the information in the output state~\eqref{eq:example_trace_final_state} can be extracted directly.

A simple way to verify this result is to perform many independent runs of the circuit, where after each run the final state of all the registers is measured and then reset to the initial state $\ket{\Omega}_g\ket{\Omega}_q\ket{\Omega}_{\tilde{q}}\ket{\Omega}_\mathcal{U}$.
One can then count the number of runs where the final state is measured to be $\ket{\Omega}_g\ket{\Omega}_q\ket{\Omega}_{\tilde{q}}\ket{\Omega}_\mathcal{U}$,
and compare against the prediction of eq.~\eqref{eq:example_trace_final_state} which is that for each run the measurement of the final state will yield $\ket{\Omega}_g\ket{\Omega}_q\ket{\Omega}_{\tilde{q}}\ket{\Omega}_\mathcal{U}$ with probability $\frac{\mathcal{C}^2}{24^2}$, where $\mathcal{C}$ was defined in eq.~\eqref{eq:example_trace_analytic_result}.\footnote{Such a verification will not yield information about the complex phase of the trace, but that can be obtained by instead implementing the well-known Hadamard test~\cite{doi:10.1098/rspa.1998.0164}.}
We note that this measurement strategy, while relatively simple to understand, is only being presented as a transparent way to verify that the state~\eqref{eq:example_trace_final_state} is correctly produced.
More advanced methods exist for examining, measuring, and exploiting quantum states. For example, Quantum Amplitude Estimation~\cite{Brassard:2000,Grinko:2019,Suzuki:2019,Nakaji:2020} could be employed in order to achieve a quadratic improvement in speed.
Alternatively, to go beyond calculating the colour factor of a single Feynman diagram, the output~(\ref{eq:example_trace_final_state}) of the circuit can be directly used as a component of a future algorithm, such as one that calculates the kinematic factors and then sums over Feynman diagrams, or one that performs Monte Carlo integration to compute cross-sections.

Although not required for this simple example Feynman diagram, let us mention that in addition to the $Q$ gate used above, we have also designed a gate $G$ which represents triple-gluon interactions. The detailed definition of $G$ will be given in sec.~\ref{sec:gluon_gate}, but for now we will simply note that, similarly to eq.~\eqref{eq:Tij_braket}, $G$ has been designed to act on 3 gluon registers ($g_1$, $g_2$, and $g_3$) and $\mathcal{U}$ such that
\begin{equation}\label{eq:fabc_braket}
G \ket{a}_{g_1}\ket{b}_{g_2}\ket{c}_{g_3}\ket{\Omega}_{\mathcal{U}} = f^{abc} \ket{a}_{g_1}\ket{b}_{g_2}\ket{c}_{g_3}\ket{\Omega}_{\mathcal{U}} + \left(\textrm{terms orthogonal to } \ket{\Omega}_\mathcal{U} \right) \,,
\end{equation}
where $f^{abc}$ are the SU(3) structure constants mentioned above.
Note that to avoid artificially distinguishing between ``emitted'' and ``emitter'' gluons, we have a separate register for each of the 3 gluons at a triple-gluon vertex and so the $G$ gate in eq.~\eqref{eq:fabc_braket} only rescales the amplitude (projected onto $\ket{\Omega}_\mathcal{U}$) by $f^{abc}$, without rotating the gluon colour states.
In contrast, the $q$ register represents an entire quark line, whose state (projected onto $\ket{\Omega}_\mathcal{U}$) is rotated by the $Q$ gate at each interaction.
Note also that we do not construct a specific gate for the four-gluon vertex since that vertex can decomposed into a linear combination of products of three-gluon vertices, each product having an independent kinematic coefficient.

This concludes our example computation of a colour trace. In sec.~\ref{sec:results}, we will generalise this to calculate the colour traces of more complicated processes.
Before that, however, we will describe the details that we have so far deferred: in sec.~\ref{sec:unitarisation_register} we will describe the purpose and functioning of the unitarisation register $\mathcal{U}$, and in sec.~\ref{sec:gates} we will present the explicit construction of the $Q$ and $G$ gates.

\subsection{Unitarisation register}
\label{sec:unitarisation_register}

To simulate perturbative QCD processes, we would like to construct quantum gates for the 8 linear operators
\begin{equation}\label{eq:linear_operator_qg}
\ket{j}_q \rightarrow \sum_i T^a_{ij} \ket{i}_q\,,
\end{equation}
and the (diagonal) linear operator
\begin{equation}\label{eq:linear_operator_ggg}
\ket{a}_{\!g_{_1}}\!\ket{b}_{\!g_{_2}}\!\ket{c}_{\!g_{_3}} \rightarrow f^{abc} \ket{a}_{\!g_{_1}}\!\ket{b}_{\!g_{_2}}\!\ket{c}_{\!g_{_3}}\,,
\end{equation}
where $q$ is a quark register and $g_{_1}, g_{_2}, g_{_3}$ are gluon registers.
However, quantum gates can only be constructed for unitary operators.
A linear operator is unitary if and only if the rows of its matrix representation are orthonormal.
The matrix representations of eqs.~\eqref{eq:linear_operator_qg} and~\eqref{eq:linear_operator_ggg} consist of rows which are mutually orthogonal but not necessarily of unit norm;\footnote{Helpfully for what follows later, the factor of $\frac{1}{2}$ in eq.~\eqref{eq:T_is_half_lambda} ensures that all rows have norm $\leq 1$.} indeed, many rows are zero.
In this section we will present a way to circumvent this problem.

Let $L$ be a linear operator acting on a Hilbert space $\mathcal{H}_1$. If $L$ is non-unitary, it cannot be implemented as a quantum circuit.
The Gell-Mann matrices in eq.~\eqref{eq:lambas} are examples of such non-unitary operations. 
However, it may still be possible to define a unitary operator $\hat{L}$ acting on a larger Hilbert space $\mathcal{H}_1 \otimes \mathcal{H_U}$ such that for some state $\ket{\Omega}_\mathcal{U} \in \mathcal{H_U}$ and for any states $\ket{\chi_1}, \ket{\chi_2} \in \mathcal{H}_1$ the following equation holds:
\begin{equation}\label{eq:linear_operator_unitarisation}
\bra{\Omega}_\mathcal{U} \braket{\chi_2|\hat{L}|\chi_1}\ket{\Omega}_\mathcal{U} = \braket{\chi_2|L|\chi_1}\,.
\end{equation}

Clearly, there are many ways to achieve eq.~\eqref{eq:linear_operator_unitarisation}, each with different advantages.
In this work, we have sought a way that firstly allows multiple independent non-unitary operations to be performed sequentially, secondly keeps the size of $\mathcal{H_U}$ small, and thirdly maintains quantum coherence without intermediate measurements so that these circuits can be used as building blocks in a higher-level algorithm.
Specifically, regardless of the complexity of the Feynman diagram, we introduce a single additional register $\mathcal{U}$, whose size is small: it contains $N_\mathcal{U} = \left \lceil \log_2(N_V + 1) \right \rceil$ qubits, where $N_V$ is the number of vertices in the Feynman diagram.
More generally, $N_V$ would be the number of non-unitary operations to be performed.
We call $\mathcal{U}$ the \emph{unitarisation register}, and denote its $2^{N_\mathcal{U}}$ basis states $\ket{k}_\mathcal{U}$ with $k \in \{0, \ldots, 2^{N_\mathcal{U}}-1 \}$.
Later in this section we will define two gates $A$ and $B(\alpha)$, where $\alpha \in \mathbb{C}$ and $|\alpha|^2 \le 1$, which are designed to act on $\mathcal{U}$ in the following way:
\begin{equation}\label{eq:BA_cycling}
B(\alpha) A \ket{k} =
\begin{cases}
\alpha \ket{0} + \sqrt{1-|\alpha|^2}\ket{1}\,, & \textrm{ if } k=0 \,,\\
\ket{k+1}\,, & \textrm{ if } 0 < k < 2^{N_\mathcal{U}} - 1\,, \\
\sqrt{1-|\alpha|^2} \ket{0} - \alpha \ket{1}\,, & \textrm{ if } k = 2^{N_\mathcal{U}} - 1\,.
\end{cases}
\end{equation}
The state $\ket{0}_\mathcal{U}$ is special and we interchangeably denote it as $\ket{\Omega}_\mathcal{U}$. Equation~\eqref{eq:BA_cycling} implies two key properties: firstly, 
\begin{equation}\label{eq:single_partial_increment}
\braket{\Omega|_\mathcal{U} B(\alpha)A|\Omega}_\mathcal{U} = \alpha\,,
\end{equation}
and secondly, we can apply the $A$ and $B$ gates\footnote{It will turn out that $B(0)=\id$, and so the separation of the $A$ and $B(\alpha)$ gates will allow the $B(\alpha)$ gate to be omitted if $\alpha=0$.} repeatedly up to $2^{N_\mathcal{U}}-1$ times and obtain
\begin{equation}
\label{eq:multiple_partial_increments}
\bra{\Omega}_\mathcal{U} \prod_{i=1}^{N_{ops}} \left\{B(\alpha_i)A\right\} \ket{\Omega}_\mathcal{U} = \prod_{i=1}^{N_{ops}} \alpha_i\,,
\end{equation}
where the number of operations $N_{ops} \leq 2^{N_\mathcal{U}}-1$.

For a given operator $L$, our general strategy for implementing an operator $\hat{L}$ which satisfies eq.~\eqref{eq:linear_operator_unitarisation} comprises two steps.
One step is to act on $\mathcal{H}_1$ with a unitary operator whose rows differ from the rows of $L$ by only a (row-dependent) normalisation.
The other step is to act on $\mathcal{H}_1 \otimes \mathcal{H_U}$ with controlled\footnote{To be explained shortly.} versions of the $A$ and $B(\alpha)$ gates in a way that, thanks to eq.~\eqref{eq:single_partial_increment}, corrects for the normalisation changes.\footnote{By applying suitable rotations, our method could be extended to apply to cases where the rows of $L$ are not mutually orthogonal, but that is beyond the scope of this article.}
Equation~\eqref{eq:linear_operator_unitarisation} follows as a direct consequence of these two steps.
To apply a sequence of non-unitary operations, we simply repeat these two steps.
In general, this will place $\mathcal{U}$ into a superposition of states, with the component proportional to $\ket{\Omega}_\mathcal{U}$ containing the information of interest due to eq.~\eqref{eq:linear_operator_unitarisation}.
Since the step acting on $\mathcal{H}_1$ does not affect $\mathcal{U}$, and the step acting on $\mathcal{H}_1 \otimes \mathcal{H_U}$ only increments the state of $\mathcal{U}$ by at most 1 according to eq.~\eqref{eq:BA_cycling} and never decrements it as long as $k<2^{N_\mathcal{U}}-1$,
it can be seen that the two steps can be repeated up to $2^{N_\mathcal{U}}-1$ times before the unitarisation register overflows.
Therefore, as desired, the required size of $\mathcal{U}$ is small: it is logarithmic in the number of sequential non-unitary operations that we wish to perform.
Furthermore, our unitarisation strategy maintains the quantum coherence of states produced by non-unitary operations, thus allowing the circuits in this paper to be used as building blocks for higher-level algorithms.
The explicit implementation of eq.~\eqref{eq:linear_operator_unitarisation} for the linear operators in eqs.~\eqref{eq:linear_operator_qg} and~\eqref{eq:linear_operator_ggg} is left to sec.~\ref{sec:gates}.
In the remainder of this section, we will introduce notation for controlled quantum gates, and then give explicit definitions for the $A$ and $B(\alpha)$ gates.

\begin{figure}[t]
\center
\begin{subfigure}{0.25\textwidth}
\includegraphics[width=\textwidth]{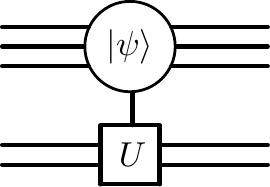}
\end{subfigure}
\hfill
\caption{\label{fig:control}%
Graphical representation of $\ket{\psi}$-controlled-$U$ gate $C_{\ket{\psi}}\left[U\right]$ defined in eq.~\eqref{eq:controlledU}. Here, $\ket{\psi}$ is any specified state of the upper register, and $U$ is any operator defined on Hilbert space of the lower register.
}
\end{figure}

For convenience, let us define some notation for controlled quantum gates, which will be used in this section and the next one.
For any quantum gate, a controlled version of it can loosely be understood as applying that gate to one register, designated as the \emph{target} register, if one or more other registers, designated as the \emph{control} registers, are in a particular specified state.
For example, as we will see in sec.~\ref{sec:quark_gate}, the $Q$ gate implements the quark-gluon interaction by applying a rotation which is targeted at the quark register, with the choice of rotation controlled by the state of the gluon register.
More precisely, let $U$ (not to be confused with $\mathcal{U}$) be any quantum gate
acting on a Hilbert space $\mathcal{H}_{\textrm{trgt}}$ and let $\ket{\psi}$ be a normalised state in another Hilbert space $\mathcal{H}_{\textrm{ctrl}}$.
Then we define the $\ket{\psi}$-controlled-$U$ gate $C_{\ket{\psi}}\left[U\right]$
acting on the Hilbert space $\mathcal{H}_{\textrm{ctrl}} \otimes \mathcal{H}_{\textrm{trgt}}$
as follows:
\begin{equation}\label{eq:controlledU}
C_{\ket{\psi}}\left[U\right] = \ket{\psi}\bra{\psi}\otimes U + \left(\id_{\mathcal{H}_{\textrm{ctrl}}}-\ket{\psi}\bra{\psi}\right)\otimes \id_{\mathcal{H}_{\textrm{trgt}}}\,.
\end{equation}
Here $\id_{\mathcal{H}_i}$ is the identity operator acting on the Hilbert space $\mathcal{H}_i$.
In the context of eq.~\eqref{eq:controlledU} we will call the qubits represented by $\mathcal{H}_{\textrm{ctrl}}$ the control qubits, and call the qubits represented by $\mathcal{H}_{\textrm{trgt}}$ the target qubits.
Equation~\eqref{eq:controlledU} implies in particular that for any state $\ket{\phi} \in \mathcal{H}_{\textrm{trgt}}$, 
\begin{equation}\label{eq:controlledU_psi_phi}
C_{\ket{\psi}}\left[U\right] \{ \ket{\psi}\otimes\ket{\phi} \} = \ket{\psi}\otimes U \ket{\phi}\,,
\end{equation}
and furthermore that given any state $\ket{\psi'} \in \mathcal{H}_{\textrm{ctrl}}$ satisfying $\braket{\psi'|\psi}=0$,
\begin{equation}\label{eq:controlledU_psiprime_phi}
C_{\ket{\psi}}\left[U\right] \{ \ket{\psi'}\otimes\ket{\phi} \} = \ket{\psi'}\otimes \ket{\phi}\,.
\end{equation}
In this article, any controlled gate $C_{\ket{\psi}}\left[U\right]$ will be depicted in the manner shown in fig.~\ref{fig:control}.
In our {\sc Qiskit} program, the function \texttt{qiskit.extensions.UnitaryGate.control} has been used to implement eq.~\eqref{eq:controlledU}.

\begin{figure}[t]
\center
\begin{subfigure}{0.7\textwidth}
\includegraphics[width=\textwidth]{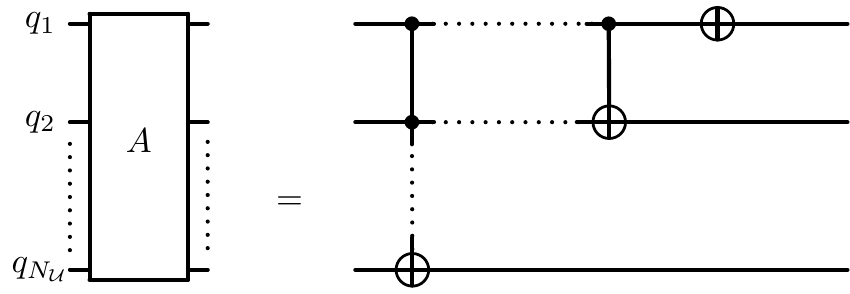}
\end{subfigure}

\caption{\label{fig:increment}%
Graphical representation of the circuit for the $A$ operator, which increments the unitarisation register $\mathcal{U}$.
Here, $\bigoplus$ represents the Pauli-$X$ gate and $\ket{q_1}, \ldots ,\ket{q_{N_\mathcal{U}}}$ are the qubits that form $\mathcal{U}$.
}
\end{figure}

Let us proceed by defining the well-known increment operator (see for example \cite{nielsen2010quantum,johnston2019programming}), which we denote $A$ and which is depicted in fig.~\ref{fig:increment}.
Formally, the gate can be defined as follows.
Let $X$ be the Pauli-$X$ single-qubit gate:
\begin{equation}
 X = \begin{pmatrix} 0 & 1 \\ 1 & 0 \end{pmatrix}\,.
\end{equation}
Let $X_n$ be the gate $C_{\ket{\underbrace{1 \ldots 1}_{n-1}}}[X]$ acting on the first $n$ qubits of the unitarisation register, with qubits $1$ through $n-1$ serving as control qubits and qubit $n$ serving as target qubit.
Let $A_1$ be the Pauli-$X$ gate acting on qubit $1$, and define recursively\footnote{One sometimes encounters~\cite{johnston2019programming} a different (but equivalent) recursive definition, which involves a controlled-$A_{n-1}$ gate, but we have chosen not to adopt it here.} for $n>1$:
\begin{equation}
A_n = A_{n-1} X_n\,.
\end{equation}
For a unitarisation register with $N_\mathcal{U}$ qubits, the increment operator of fig.~\ref{fig:increment} can now be formally defined as $A = A_{N_\mathcal{U}}$.
Interpreting each basis state of the unitarisation register as a binary representation of a number $\ket{k = \sum_{i=1}^{N_\mathcal{U}} u_i 2^{i-1}}$, where $u_n\in\{0,1\}$ is the state of the $n^{\textrm{th}}$ qubit, it can be verified that
\begin{equation}\label{eq:increment_modular_definition}
A\ket{k}_\mathcal{U} = \ket{k+1\textrm{ (mod } 2^{N_\mathcal{U}}\textrm{)}}_\mathcal{U}\,, \quad \forall k\,.
\end{equation}
In this article, $A$ should be assumed to always act on the unitarisation register.

Next let us define a single-qubit gate
\begin{equation}\label{eq:single_qubit_partial_decrement}
  B_1(\alpha) =
  \begin{pmatrix}
    \sqrt{1-|\alpha|^2} & \alpha \\ -\alpha & \sqrt{1-|\alpha|^2}
  \end{pmatrix}\,,
\end{equation}
which should be understood to always act on qubit 1 of the unitarisation register.

\begin{figure}
\center
\begin{subfigure}{0.50\textwidth}
\includegraphics[width=\textwidth]{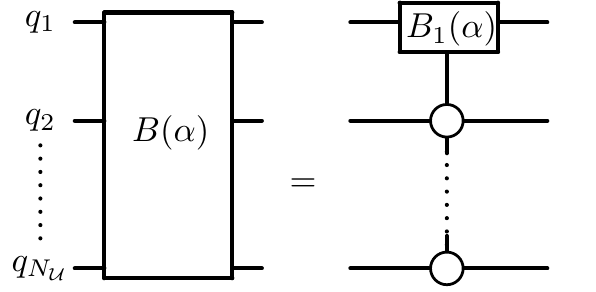}
\end{subfigure}
\caption{\label{fig:decrement}%
Graphical representation of the circuit for the partial-decrement operator $B(\alpha)$ defined formally in eq.~\eqref{eq:partial_decrement_definition}.
}
\end{figure}

We now define a partial-decrement operator $B(\alpha)$ as follows:
\begin{equation}\label{eq:partial_decrement_definition}
B(\alpha) = C_{\ket{\underbrace{0 \ldots 0}_{N^{}_{\mathcal{U}}\textrm{ --- } 1}}}[B_1(\alpha)]\,,
\end{equation}
which should be understood to act on the unitarisation register with qubits $2$ through $N_\mathcal{U}$ serving as control qubits and qubit $1$ serving as target qubit, as shown in fig.~\ref{fig:decrement}.
The desired behaviour of the $A$ and $B(\alpha)$ gates shown in eq.~\eqref{eq:BA_cycling} follows immediately from eqs.~(\ref{eq:increment_modular_definition}--\ref{eq:partial_decrement_definition}).

This concludes our explanation of the unitarisation register $\mathcal{U}$ and the $A$ and $B(\alpha)$ gates. We will now proceed to use them to implement eq.~\eqref{eq:linear_operator_unitarisation} for the quark-gluon~\eqref{eq:linear_operator_qg} and triple-gluon~\eqref{eq:linear_operator_ggg} interactions.

\subsection{Construction of interaction gates}
\label{sec:gates}

In this section, we will give details of the construction of the $Q$ and $G$ gates which simulate quark-gluon and triple-gluon interactions, respectively.
As mentioned earlier, 2-qubit registers are used to represent the $N_{\rm c} = 3$ colour states of a quark as $\ket{00}$, $\ket{01}$, and $\ket{10}$. 
Note that the $\ket{11}$ state of each quark register is unused: all operators $\mathcal{O}$ acting on a quark register
satisfy $\braket{11|\mathcal{O}|11}=1$, which means the last row and last column of the $4 \times 4$ matrix representing $\mathcal{O}$ are both $(0\quad 0 \quad 0 \quad 1)$,
ensuring that the register never enters the $\ket{11}$ state.
For brevity we will omit the fourth row and fourth column when representing these operators as matrices.

\subsubsection{Quark-gluon interaction gate $Q$}
\label{sec:quark_gate}

\begin{figure}[t]
\center
\begin{subfigure}{0.55\textwidth}
\includegraphics[width=\textwidth]{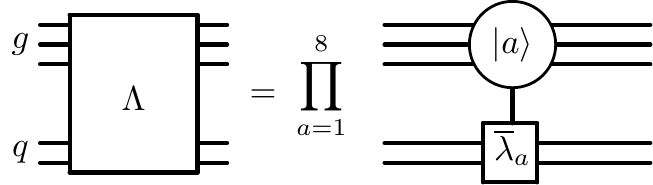}
\end{subfigure}
\caption{\label{fig:L_gate}%
Graphical representation of the circuit of the $\Lambda$ gate defined in eq.~\eqref{eq:L}.
}
\end{figure}

We wish to construct a gate $Q$ which will implement the quark-gluon interaction.
This interaction is described by the non-unitary operator shown in eq.~\eqref{eq:linear_operator_qg}, and we therefore wish unitarise it---see eq.~\eqref{eq:linear_operator_unitarisation}---by constructing a suitable $\hat{L}$ for it.
Following the general strategy explained in sec.~\ref{sec:unitarisation_register}, we start by defining the following unitary matrices $\overline{\lambda}_a$:
\begin{equation}\label{eq:lambda_bars}
\begin{gathered}
\overline{\lambda}_1 = \matthree {0}{1}{0}{1}{0}{0}{0}{0}{1}\,,\quad
\overline{\lambda}_2 = \matthree {0}{-i}{0}{i}{0}{0}{0}{0}{1}\,,\quad
\overline{\lambda}_3 = \matthree {1}{0}{0}{0}{-1}{0}{0}{0}{1},\\[1ex]
\overline{\lambda}_4 = \matthree {0}{0}{1}{0}{1}{0}{1}{0}{0}\,,\quad
\overline{\lambda}_5 = \matthree {0}{0}{-i}{0}{1}{0}{i}{0}{0}\,,\quad
\overline{\lambda}_6 = \matthree {1}{0}{0}{0}{0}{1}{0}{1}{0},\\[1ex]
\overline{\lambda}_7 = \matthree {1}{0}{0}{0}{0}{-i}{0}{i}{0}\,,\quad
\overline{\lambda}_8 = \matthree {1}{0}{0}{0}{1}{0}{0}{0}{1} \,.
\end{gathered}
\end{equation}
It can be observed that these matrices are similar to the Gell-Mann matrices eq.~\eqref{eq:lambas} but have been adjusted to make them unitary, and therefore implementable as quantum gates acting on a 3-state quark register.
We combine the $\overline{\lambda}_a$ gates into a new gate $\Lambda$ acting on a gluon register $g$ and quark register $q$ in the following way
\begin{equation}
\label{eq:L}
\Lambda = \left[\prod_{a=1}^8 C_{\ket{a}_g}\left[\overline{\lambda}_a\right]\right]\,,
\end{equation}
where $g$ serves as control register and $q$ serves as target register, as shown in fig.~\ref{fig:L_gate}.
Thus, depending on the colour $a$ of the gluon register, $\Lambda$ applies the gate $\overline{\lambda}_a$ to the quark register.

\begin{figure}[t]
\center
\begin{subfigure}{0.8\textwidth}
\includegraphics[width=\textwidth]{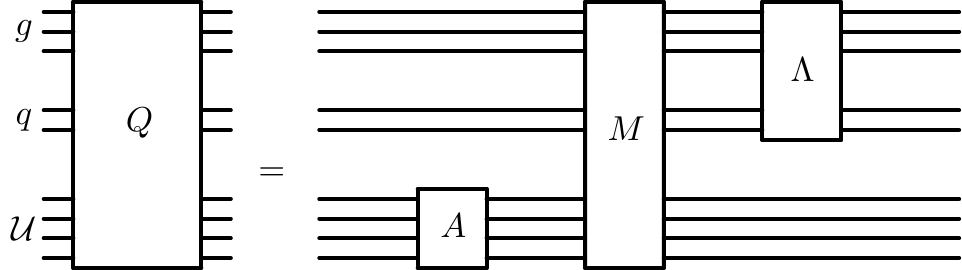}
\end{subfigure}
\caption{\label{fig:quark_gates}%
Graphical representation of the circuit of the quark gate $Q$ defined in eq.~\eqref{eq:Q}.
}
\end{figure}
\begin{figure}[t]
\center
\begin{subfigure}{0.55\textwidth}
\includegraphics[width=\textwidth]{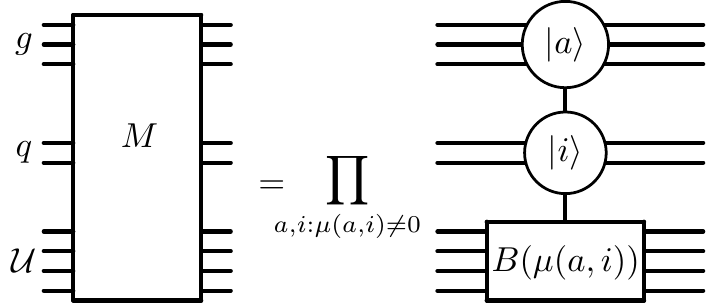}
\end{subfigure}
\caption{\label{fig:M_gate}%
Graphical representation of the circuit of the $M$ operator defined in eq.~\eqref{eq:M}.
}
\end{figure}

We proceed to construct the quark-gluon interaction gate $Q$, which is based on $\Lambda$ but uses $A$ and $B$ gates to account for the differences between the matrices $\lambda_a$ and $\overline{\lambda}_a$, as well as for the factor of $\frac{1}{2}$ in eq.~\eqref{eq:T_is_half_lambda}.
In brief, the $Q$ gate will increment $\mathcal{U}$ using the $A$ gate and then conditionally decrement it again using $B$ gates, before finally applying the $\Lambda$ gate to the state thus produced.
More formally, as depicted in fig.~\ref{fig:quark_gates}, the $Q$ gate is defined to act on the state $\ket{\Psi}_{g \, \otimes \, q \, \otimes \, \mathcal{U}}$ of any gluon register $g$, any quark register $q$, and the unitarisation-register $\mathcal{U}$ in the following way:
\begin{align}
\label{eq:Q}
Q\ket{\Psi}_{g \, \otimes \, q \, \otimes \, \mathcal{U}} = \left(\Lambda \otimes \id_\mathcal{U} \right) M (\id_g \otimes \id_q \otimes A) \ket{\Psi}_{g \, \otimes \, q \, \otimes \, \mathcal{U}}\,,
\end{align}
where
\begin{equation}
\label{eq:M}
M = \prod_{a,i \textrm{ : } \mu(a,i)\neq 0} C_{\ket{a}_g\ket{i}_q}\left[B\left(\mu(a,i)\right)\right]\,,
\end{equation}
and
\begin{equation}\label{eq:scaling_lambdabars}
\mu(a,i) =
\begin{cases}
\frac{1}{2}\,, & \textrm{ if } \left(\overline{\lambda}_a\right)_{ij} - \left(\lambda_a\right)_{ij} = 0\,, \quad \forall j\,,\\
\frac{1}{2\sqrt{3}}\,, & \textrm{ if } a = 8 \,,\textrm{ and } i \in \{1,2\}\,, \\
\frac{-1}{\sqrt{3}}\,, & \textrm{ if } a = 8\,, \textrm{ and } i = 3\,, \\
0\,, & \textrm{ otherwise.}
\end{cases}
\end{equation}
A graphical representation of the operator $M$ is shown in fig.~\ref{fig:M_gate}.

The definitions of $\overline{\lambda}_a$ and $\mu(a,i)$ in eqs.~\eqref{eq:lambda_bars} and \eqref{eq:scaling_lambdabars} have been chosen so that the following equation holds
\begin{equation}
\mu(a,i) \overline{\lambda}_a \ket{i} = \dfrac{1}{2}\lambda_a \ket{i}\,,
\end{equation}
where the $\lambda_a$ are defined in eq.~\eqref{eq:lambas} and the factor of $\frac{1}{2}$ originates from eq.~\eqref{eq:T_is_half_lambda}.
Note that since the control states in eq.~\eqref{eq:M} are mutually orthogonal, applying $M$ does not decrement the state of $\mathcal{U}$ by more than 1.
By using the properties of $\mathcal{U}$ in eq.~\eqref{eq:single_partial_increment} and the definitions of $Q$, $M$, and $\Lambda$ in eqs.~(\ref{eq:L}--\ref{eq:M}), it can be seen that if $\ket{\psi_1} = \ket{a}_{g}\ket{k}_{q}\ket{\Omega}_\mathcal{U}$ and $\ket{\psi_2} = \ket{b}_{g}\ket{l}_{q}\ket{\Omega}_\mathcal{U}$ then
\begin{equation}
\braket{\psi_2|Q|\psi_1} =
\braket{b|a} \braket{l | \tfrac{1}{2} \lambda_a |k}\,.
\end{equation}
The desired property of $Q$ shown in eq.~\eqref{eq:Tij_braket} follows immediately from this.
Thus, a sequence of emissions and absorptions of gluons by a quark line can be simulated on a quantum computer by chaining a corresponding sequence of $Q$ gates.

\subsubsection{Triple-gluon interaction gate $G$}
\label{sec:gluon_gate}

We shall now proceed to the description of a quantum gate for the triple-gluon interaction.
This interaction is described by the non-unitary operator shown in eq.~\eqref{eq:linear_operator_ggg}, and so we wish to unitarise it---see eq.~\eqref{eq:linear_operator_unitarisation}---by finding a suitable $\hat{L}$ for it, following the general strategy explained in sec.~\ref{sec:unitarisation_register}.
In contrast to the quark-gluon operator~\eqref{eq:linear_operator_qg}, the triple-gluon operator~\eqref{eq:linear_operator_ggg} is diagonal and so its corresponding $\hat{L}$ operator, which we call $G$, can be constructed using only $A$ and controlled $B(\alpha)$ gates.

\begin{figure}[t]
\center
\begin{subfigure}{0.60\textwidth}
\includegraphics[width=\textwidth]{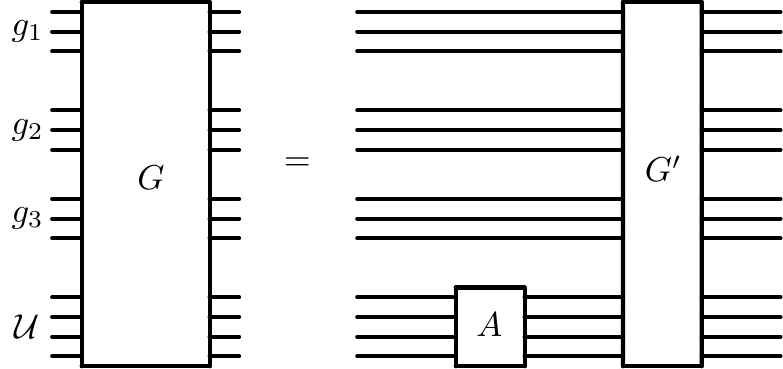}
\end{subfigure}
\caption{\label{fig:G_gate}%
Graphical representation of the gluon gate $G$ defined in eq.~\eqref{eq:G}.
}
\end{figure}

\begin{figure}[t]
\center
\begin{subfigure}{0.55\textwidth}
\includegraphics[width=\textwidth]{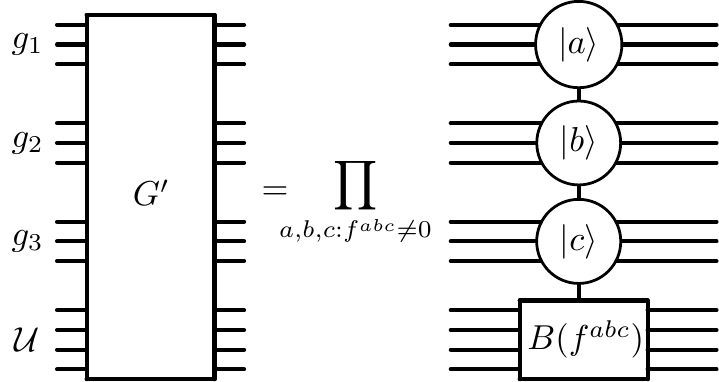}
\end{subfigure}
\hfill
\caption{\label{fig:Gp_gate}%
Graphical representation of the circuit of the $G'$ gate defined in eq.~\eqref{eq:Gp}.
}
\end{figure}

We define the gate $G$ acting on any state $\ket{\Psi}_{\! g_{_1} \otimes\, g_{_2} \otimes\, g_{_3} \otimes\, \mathcal{U}}$ of gluon registers $g_1$, $g_2$, and $g_3$, and the unitarisation register $\mathcal{U}$ in the following way:
\begin{equation}
\label{eq:G}
G \ket{\Psi}_{\! g_{_1} \otimes\, g_{_2} \otimes\, g_{_3} \otimes\, \mathcal{U}} = G' A \ket{\Psi}_{\! g_{_1} \otimes\, g_{_2} \otimes\, g_{_3} \otimes\, \mathcal{U}}  \,,
\end{equation}
where
\begin{equation}
\label{eq:Gp}
G' \ket{\Psi}_{\! g_{_1} \otimes\, g_{_2} \otimes\, g_{_3} \otimes\, \mathcal{U}} = \left( \prod_{a,b,c \textrm{ : } f^{abc}\neq 0} C_{\ket{a}\ket{b}\ket{c}}\left[B\left(f^{abc}\right)\right] \right) \ket{\Psi}_{\! g_{_1} \otimes\, g_{_2} \otimes\, g_{_3} \otimes\, \mathcal{U}}  \,.
\end{equation}
The gates $G$ and $G'$ are illustrated in figs.~\ref{fig:G_gate} and \ref{fig:Gp_gate}, respectively.

\begin{tolerant}{3000}
One sees that in eq.~\eqref{eq:Gp}, by definition~(\ref{eq:controlledU}) of the control gate, each factor $C_{\ket{a}\ket{b}\ket{c}}\left[B\left(f^{abc}\right)\right]$ applies $B(f^{abc})$ to the unitarisation register if the three gluons have colours $a,b,c$ respectively, and leaves the unitarisation register unchanged if the gluons are in a state orthogonal to $\ket{a}_{g_{_1}}\ket{b}_{g_{_2}}\ket{c}_{g_{_3}}$.
As was seen for the $M$ gate in the previous section, applying the $G'$ gate does not decrement the state of $\mathcal{U}$ by more than 1, since the control states in eq.~\eqref{eq:Gp} are mutually orthogonal.
Since $B(0) = \id$ and $C_{\ket{a}\ket{b}\ket{c}}\left[\id\right] = \id$, the product in eq.~\eqref{eq:Gp} does not need to include any cases where $f^{abc}=0$.
By using the property shown in eq.~\eqref{eq:single_partial_increment} and the definitions of $G$ and $G'$ in eqs.~\eqref{eq:G} and \eqref{eq:Gp}, it can be verified that eq.~\eqref{eq:fabc_braket} indeed holds.
\end{tolerant}
Hence, a triple-gluon interaction can be implemented by applying the gate $G$ to the corresponding gluon registers. It can be observed that the gates $G$ and $Q$ do not rotate the states of the gluon registers and so in diagrams where several triple-gluon interactions are present, the corresponding $G$ gates can be applied in any order.

\section{Results}
\label{sec:results}

In this section, the method introduced in sec.~\ref{sec:colour_traces} will be generalised to simulate---and calculate colour factors for---arbitrary Feynman diagrams.

Let $n_g$ be the number of gluons in the diagram and let $n_q$ be the number of quark lines in the diagram.
The quantum circuit to be constructed will contain gluon registers, quark registers, and a unitarisation register.
There are $n_g$ gluon registers, each with 3 qubits.
There are $n_q$ pairs of quark registers, each pair comprising 2 registers labelled $q$ and $\tilde{q}$, with 2 qubits per register.
The unitarisation register has $N_\mathcal{U} = \left \lceil \log_2(N_V + 1) \right \rceil$ qubits, where $N_V$ is the number of vertices in the Feynman diagram.
The procedure for calculating colour factors is as follows:
\begin{enumerate}
 \item Initialise each register $r$ into the state $\ket{\Omega}_r$.
 \item Apply $R_g$, as in eq.~(\ref{eq:Rg_effect}), to each gluon register separately.
 \item For each quark line, apply $R_q$ to the corresponding pair of quark registers, $q$ and $\tilde{q}$, as in eq.~(\ref{eq:Rq_effect}).
 \item For each quark-gluon interaction vertex, apply a $Q$ gate to the quark register $q$ and gluon register $g$ that correspond to the quark and gluon at that vertex. The corresponding $\tilde{q}$ register does not participate here.
 \item For each triple-gluon interaction, apply a $G$ gate to the 3 corresponding gluon registers.
 \item Apply an $R_g^{-1}$ gate to each gluon register.
 \item For each quark line, apply an $R_q^{-1}$ gate to the corresponding pair of quark registers, $q$ and~$\tilde{q}$.
\end{enumerate}
In the above procedure, steps 4 and 5 simulate the evolution of the colour states of the particles in the Feynman diagram, while the remaining steps serve to perform the trace over the colours.
Note that the $Q$ gates corresponding to a given quark line must be applied in the order in which the corresponding interactions appear on that quark line in the Feynman diagram.
Apart from this, there is no restriction on the ordering of the $Q$ and $G$ gates.

\begin{table}[t!]
\centering
\caption{\label{tab:checks}
Colour factors for example Feynman diagrams.
The first column depicts the Feynman diagrams, with indices on external legs indicating identical colours.
The central column states the analytical result for the colour factor.
The last column displays the numerical result for each colour factor obtained using $100$ million runs of the simulated quantum circuit, along with the associated statistical uncertainty.
}
\begin{tabular}{c|c|c}
Diagram & Analytical & Numerical \\
\midrule
\thead{\includegraphics[width=0.3\textwidth]{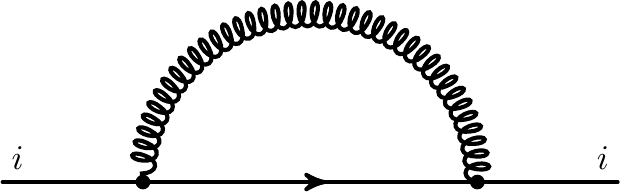}} & $C_F N = 4$ & $3.9988 \pm 0.0012$ 
\\
\midrule
\thead{\includegraphics[width=0.3\textwidth]{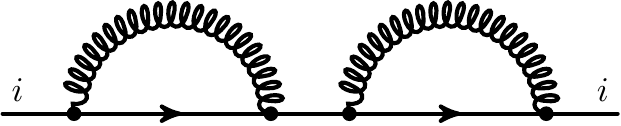}} & ${C_F}^2 N = \frac{16}{3}$ & $5.331 \pm 0.010$ 
\\
\midrule
\thead{\includegraphics[width=0.3\textwidth]{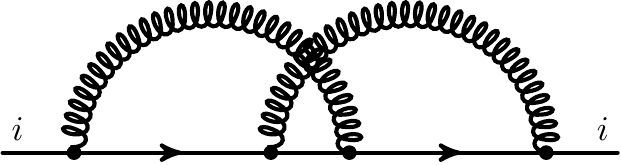}} & $\frac{C_F}{2} = \frac23$ & $0.673 \pm 0.010$ 
\\
\midrule
\thead{\includegraphics[width=0.3\textwidth]{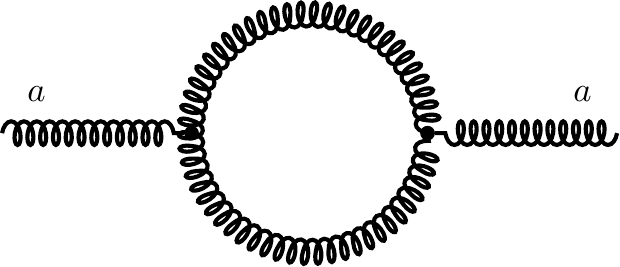}} & $N(N^2-1) = 24$ & $23.95 \pm 0.03$ 
\\
\midrule
\thead{\includegraphics[width=0.3\textwidth]{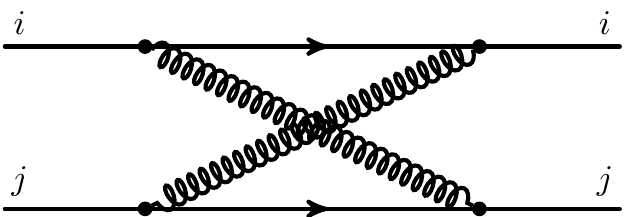}} & $\frac{(N^2-1)}{4} = 2$ & $2.00 \pm 0.03$ 
\\
\midrule
\thead{\includegraphics[width=0.3\textwidth]{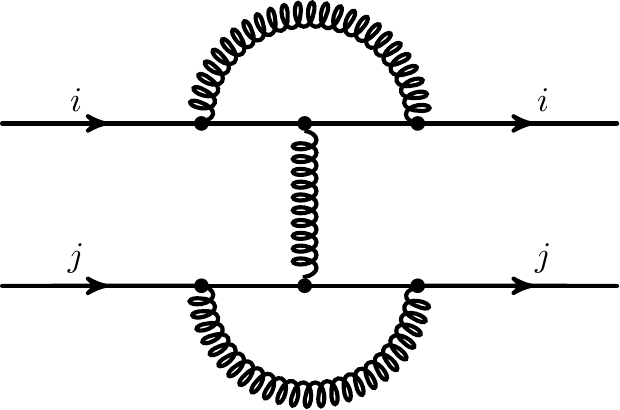}} & $0$ & $0.0^{+0.5}_{-0.0}$ 
\\
\midrule
\thead{\includegraphics[width=0.3\textwidth]{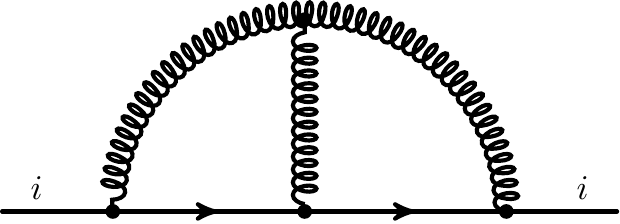}} & $\frac{C_F N^2}{2} = 6$ & $5.92 \pm 0.08$ 
\\
\end{tabular}
\end{table}

Analogously to the result in sec.~\ref{sec:colour_traces}, it follows from the Feynman rules and from eqs.~\eqref{eq:Tij_braket}, ~\eqref{eq:fabc_braket}, and~\eqref{eq:multiple_partial_increments} that after step 7 the colour factor $\mathcal{C}$ of the Feynman diagram will be found encoded in the final state of the quantum computer, which is
\begin{equation}\label{eq:final_state}
\frac{1}{\mathcal{N}} \mathcal{C} \ket{\Omega}_{all} + \left(\textrm{terms orthogonal to} \ket{\Omega}_{all}\right) \,,
\end{equation}
where $\mathcal{N} = N_c^{n_q} \left(N_c^2-1\right)^{n_g}$ and
\begin{equation}
\ket{\Omega}_{all} = \left( \prod_{m=1}^{n_g} \ket{\Omega}_{g_m} \right) \left( \prod_{l=1}^{n_q} \ket{\Omega}_{q_l}\ket{\Omega}_{\tilde{q}_l} \right) \ket{\Omega}_{\mathcal{U}}\,.
\end{equation}

As explained in sec.~\ref{sec:colour_traces}, the result in eq.~\eqref{eq:final_state} can be verified by repeatedly running the circuit and counting the number of times it is measured\footnote{The measurements are performed in the computational basis.} to be in the state $\ket{\Omega}_{all}$, optionally in conjunction with Hadamard testing (to retain phase information) and Quantum Amplitude Estimation (to obtain a quadratic speedup). Alternatively, the output~(\ref{eq:final_state}) of the circuit can be directly used as a component of a future algorithm, such as one that calculates the kinematic factors and then sums over Feynman diagrams, or one that performs Monte Carlo integration to compute cross-sections.
Since the output~\eqref{eq:final_state} is a quantum state, once combined with the corresponding kinematic factor it would be well-suited to future work which calculates the quantum interference of multiple diagrams, possibly by implementing a quantum-computing equivalent of the recursive algorithms~\cite{Berends:1987me} that are widely used in modern classical calculations.

To validate the methods and circuits described in this article, we have implemented them in a program named \textsc{QColour} written in \textsc{Python} using the IBM \textsc{Qiskit} package.\footnote{For this implementation, we have used {\sc Qiskit} version $0.36.1$.}
Using \textsc{QColour} we have built quantum circuits to simulate several Feynman diagrams shown in Table~\ref{tab:checks}.
For each Feynman diagram, the corresponding circuit was run $10^8$ times on a simulated noiseless quantum computer and the number $N_\Omega$ of times the output state was measured to be $\ket{\Omega}_{all}$ was counted.
It follows from eq.~\eqref{eq:final_state} that $N_\Omega$ is binomially distributed as $N_\Omega \sim B\left(10^8, \left(\frac{\mathcal{C}}{\mathcal{N}}\right)^2\right)$.
We therefore infer the absolute value of the colour factor to be
\begin{equation}\label{eq:inferred_colour_factor}
|\mathcal{C}| = \mathcal{N} \sqrt{\frac{N_\Omega}{10^8}}\,,
\end{equation}
with a statistical uncertainty that can be estimated using the Wilson score interval~\cite{doi:10.1080/01621459.1927.10502953}.
As can be seen from Table~\ref{tab:checks}, the colour factors obtained using the simulated quantum circuits are fully consistent with the colour factors calculated analytically.

It may be observed that the fractional uncertainty in the inferred colour factors increases with the complexity of the diagram, but we emphasise again that the measurement strategy employed here is only intended as a transparent way to verify that the circuits correctly produce the state~\eqref{eq:final_state}.
As mentioned above and in sec.~\ref{sec:colour_traces}, more sophisticated strategies can be employed to examine, measure, and exploit this quantum state, and the state can furthermore be used in a higher-level algorithm rather than being immediately measured.
Therefore, while more complicated scattering processes are always likely to have higher computational costs (as in classical Monte Carlo calculations), the examples in Table~\ref{tab:checks} should not be taken as providing a conclusive indication of the scaling rate.

\section{Conclusion}
\label{sec:conclusion}

The simulation of quantum systems is a flagship application of quantum computers, with expectations for polynomial or exponential speed-ups over classical computers.
In this article, first steps were taken towards a quantum simulation of generic perturbative QCD processes.
In particular, quantum circuits were designed to simulate the colour parts of the interactions of quarks and gluons.
In order to do so, the concept of a \emph{unitarisation register} was devised to enable a unitary quantum-circuit implementation of the non-unitary Gell-Mann matrices $\lambda_a$ and structure constants $f^{abc}$ that describe the interaction vertices in Feynman diagrams.
It was shown that these quantum circuits can be used to simulate the colour parts of arbitrary Feynman diagrams.
Furthermore, these circuits were implemented on a simulated noiseless quantum computer using the \textsc{Qiskit} framework, and colour factors were hence calculated for various examples of Feynman diagrams.
It is to be emphasised that besides enabling the calculation of colour factors, the quantum circuits presented in this work can in the future be directly used as components of a full quantum simulation of scattering amplitudes.

The work presented here opens several directions for future exploration.
Following the present simulation of the colour parts of Feynman diagrams, a natural extension of this work is the simulation of the kinematic parts, since the unitarisation register devised here would also be particularly useful there.
Simulating the kinematic parts will require appropriate ways to handle the much larger Hilbert spaces resulting from the continuous nature of kinematic variables.
This is likely to require many more qubits than the colour parts do, but since colliders like the LHC probe energy scales that are several orders of magnitude higher than $\Lambda_{\textrm{QCD}}$, we expect that the number of qubits required for simulating the kinematic parts can still be competitive against the requirements of lattice-based quantum-computer simulations of the same processes.
Another natural extension is to explore the quantum-coherent interference of the contributions from multiple Feynman diagrams, a task to which quantum computers are naturally well-suited.
Furthermore, it would also be interesting to explore the application of the quantum circuits from this work to perform calculations with full quantum correlations for high-multiplicity processes that are currently described using parton showers.
Finally, although the present work is aimed at error-corrected quantum computers that are envisaged for the medium term, it would meanwhile be interesting to test these circuits against the noise characteristics of specific near-term hardware devices, and then perform custom adaptations (in hardware and software) to mitigate against the noise and its effects.
In the long term, all these aspects can be combined with quantum algorithms known to have quadratic (or better) speedups, such as quantum Monte Carlo simulations, and then implemented on future physical quantum computers.
This would provide significant improvements in the speed and possibly the reach of perturbative QCD calculations.

\section*{Acknowledgments}

The authors are grateful to Fabrizio Caola, Stefano Gogioso, Michele Grossi, and Joseph Tooby-Smith for helpful discussions.
The authors thank the Funivia ristorante, Laveno-Mombello, Italy for providing an inspiring atmosphere which initiated this work.

\paragraph{Funding information}
The research of H.C.\ is supported by ERC Starting Grant 804394 \textsc{hip}QCD.
M.P.\ acknowledges support by the German Research Foundation (DFG) through the Research Training Group RTG2044.

\appendix
\numberwithin{equation}{section}
\label{sec:appendix}

\section{Miscellaneous gates}\label{appendix:misc_gates}
In this Appendix we give explicit definitions for the gates $R_g$ and $R_q$. Both of these gates are used in sec.~\ref{sec:colour_traces} and sec.~\ref{sec:results}. We also provide a circuit diagram for the $\Lambda$ gate that was defined in eq.~\eqref{eq:L}.

\begin{figure}[t]
\center
\begin{subfigure}{0.47\textwidth}
\includegraphics[width=\textwidth]{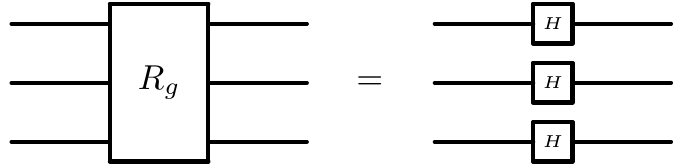}
\end{subfigure}
\caption{\label{fig:Rg_gate}%
Graphical representation of the circuit of the $R_g$ gate.
}
\end{figure}

The $R_g$ gate is composed of a Hadamard gate $H$ acting on each qubit of the gluon register.
Since $H$ satisfies $H^{-1} = H$, it follows that $R_g^{-1} = R_g$.
Its graphical representation is provided in fig.~\ref{fig:Rg_gate}.

\pagebreak

\begin{figure}[t]
\center
        \begin{subfigure}{0.47\textwidth}
        \includegraphics[width=\textwidth]{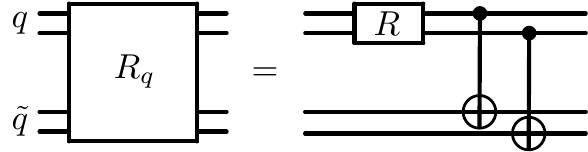}
        \end{subfigure}
        \caption{\label{fig:Rq_gate}%
                Graphical representation of the circuit of the $R_q$ gate.
                }
\end{figure}

The $R_q$ gate is defined by the circuit shown in fig.~\ref{fig:Rq_gate} and is composed of two controlled-$X$ gates and a 2-qubit gate $R$ defined as follows:
\begin{equation}
R =
\begin{pmatrix}
\sqrt{\frac{1}{3}}	&	\sqrt{\frac{1}{2}}	&	\sqrt{\frac{1}{6}}	&	0	\\
\sqrt{\frac{1}{3}}	&	-\sqrt{\frac{1}{2}}	&	\sqrt{\frac{1}{6}}	&	0	\\
\sqrt{\frac{1}{3}}	&	0	&	\sqrt{\frac{2}{3}}	&	0	\\
0	&	0	&	0	&	1
\end{pmatrix}\,.
\end{equation}
The inverse $R_q^{-1}$ of $R_q$ is easily constructed by reversing the order of the 3 gates in fig.~\ref{fig:Rq_gate} and replacing $R$ by its transpose $R^T$.

\begin{figure}[h!]
\center
        \begin{subfigure}{0.67\textwidth}
        \includegraphics[width=\textwidth]{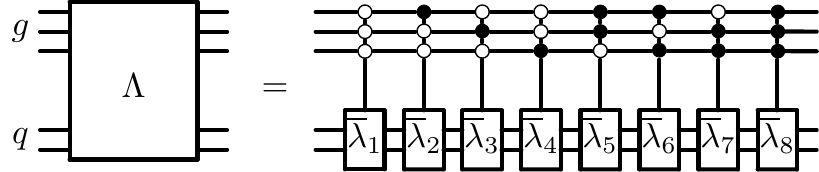}
        \end{subfigure}
        \caption{\label{fig:L_gate_detailed}%
                Explicit graphical representation of the circuit of the $\Lambda$ gate defined in eq.~\eqref{eq:L}. White and black circles represent controlled operations with control states $\ket{0}$ and $\ket{1}$, respectively.
                }
\end{figure}

Finally, the explicit form of the $\Lambda$ gate in terms of basic gates can be found in fig.~\ref{fig:L_gate_detailed}.
The matrices $\overline{\lambda}_a$ are given in eq.~\eqref{eq:lambda_bars}, and should be understood to be $4 \times 4$ dimensional as explained at the start of sec.~\ref{sec:gates}.
The $M$ gate is constructed in a similar fashion, and comprises 17 controlled gates corresponding to the 17 combinations of $a$ and $i$ for which $\mu(a,i)$ is non-zero according to eq.~\eqref{eq:scaling_lambdabars}.
However, since $\mu$ takes one of only 4 different values, we expect that $M$ could be constructed with fewer controlled operations by choosing a suitable encoding of $a$ and $i$ into qubits.

\end{document}